# 4D-Flow MRI Pressure Estimation Using Velocity Measurement-Error based Weighted Least-Squares

Jiacheng Zhang, Melissa C. Brindise, Sean Rothenberger, Susanne Schnell, Michael Markl, David Saloner, Vitaliy L. Rayz, and Pavlos P. Vlachos

*Abstract*—This work introduces a 4D-flow magnetic resonance imaging (MRI) pressure reconstruction method which employs weighted least-squares (WLS) for pressure integration. Pressure gradients are calculated from the velocity fields, and velocity errors are estimated from the velocity divergence for incompressible flow. Pressure gradient errors are estimated by propagating the velocity errors through Navier-Stokes momentum equation. A weight matrix is generated based on the pressure gradient errors, then employed for pressure reconstruction. The pressure reconstruction method was demonstrated and analyzed using synthetic velocity fields as well as Poiseuille flow measured using *in vitro* 4D-flow MRI. Performance of the proposed WLS method was compared to the method of solving the pressure Poisson equation which has been the primary method used in the previous studies. Error analysis indicated that the proposed method is more robust to velocity measurement errors. Improvement on pressure results was found to be more significant for the cases with spatially-varying velocity error level, with reductions in error ranging from 50% to over 200%. Finally, the method was applied to flow in a patient-specific cerebral aneurysm. Validation was performed with *in vitro* flow data collected using Particle Tracking Velocimetry (PTV) and Shake the Box (STB) method, and *in vivo* flow measurement obtained using 4D-flow MRI. Pressure calculated by WLS, as opposed to the Poisson equation, was more consistent with the flow structures and showed better agreement between the *in vivo* and *in vitro* data. These results suggest the utility of WLS method to obtain reliable pressure field from clinical flow measurement data.

## I. INTRODUCTION

Pressure measured from the cardiovascular system is widely used to diagnose disease. Many pressure-based clinical biomarkers, such as pulmonary wedge pressure [1], are single point measurements typically acquired by placing a pressure catheter in the region of interest [3]. However, this approach is invasive and still only provides a point measurement. Conversely, a spatial pressure distribution can provide a more complete view of the hemodynamics in the cardiovascular system. For example, the pressure distribution on a vascular wall can help predict the progression or rupture of cerebral aneurysms [2]. Further, such pressure distributions can be obtained noninvasively. One such noninvasive approach includes estimating the pressure difference from Doppler echocardiography and is typically employed for evaluating intra-ventricular pressure difference [4], [5]. However, conventional Doppler Ultrasound only measures only one component of the velocity which limits the accuracy of the estimated pressure difference. Pressure fields can also be obtained using computational fluid dynamics (CFD) simulations, but fidelity of the simulation depends on the accuracy of segmentation and boundary conditions on the walls and flow waveforms. High-resolution CFD simulations are also computationally expensive. Pressure reconstruction methods have become increasingly of interest with the development of flow measurement techniques such as 4D-flow magnetic resonance imaging (MRI) which measures time-resolved velocity fields. However, the limitations on dynamic range, signal-to-noise ratio, and spatiotemporal resolution inherent to velocity measurements using 4D-flow MRI [6] result in unreliable pressure fields. Thus, a robust algorithm is needed to accurately reconstruct the pressure field from 4D-flow MRI.

Several algorithms have been proposed to evaluate the pressure field from measured flow data. Most algorithms contain two major steps. The pressure gradient field or divergence of the pressure gradients is first calculated from the velocity fields. These are then spatially integrated to obtain the instantaneous pressure fields.

For blood flow, the pressure gradient can be calculated using the incompressible Navier-Stokes momentum equation in the following form:

$$\nabla \boldsymbol{p} = -\rho \frac{D\boldsymbol{u}}{Dt} + \mu \nabla^2 \boldsymbol{u}, \qquad (1)$$

This work is funded by the NIH R21 NS106696 grant.

J. Zhang, M. C. Brindise, and P. P. Vlachos are with the School of Mechanical Engineering, Purdue Unversity, West Lafayette, IN 47907 USA.

P. P. Vlachos is also with the Weldon School of Biomedical Engineering, Purdue Unversity, West Lafayette, IN 47907 USA (e-mail: pvlachos@purdue.edu).

S. Rothenberger and V. L. Rayz are with the Weldon School of Biomedical Engineering, Purdue Unversity, West Lafayette, IN 47907 USA.

S, Schnell and M. Markl are with the Feinberg School of Medicine, Northwestern University, Chicago, IL 60611 USA.

M. Markl is also with the McCormick School of Engineering, Northwestern University, Evanston, IL 60208, USA.

D. Saloner is with the Department of Radiology and Biomedical Imaging, University of California San Francisco, CA 94143 USA.

where $\boldsymbol{p}$ is pressure (Pa), $\nabla$ is the spatial gradient operator such that $\nabla \boldsymbol{p}$ is the pressure gradient (Pa/m), $\rho$ and $\mu$ are the density (kg/m³) and dynamic viscosity (Pa·s) of the fluid, respectively, $\boldsymbol{u}$ is the velocity (m/s), and $t$ is time (s). $\mu \nabla^2 \boldsymbol{u}$ represents viscous diffusion and $\frac{D\boldsymbol{u}}{Dt}$ represents the material acceleration (m/s²). For gridded velocity data, the material acceleration can be evaluated as:

$$\frac{D\boldsymbol{u}}{Dt} = \frac{\partial \boldsymbol{u}}{\partial t} + (\boldsymbol{u} \cdot \nabla)\boldsymbol{u}, \qquad (2)$$

[7]–[11]. For flow measurements using particle trajectories, the material acceleration can be directly determined from particle tracks using $\frac{d^2 \boldsymbol{x}_p}{dt^2}$ where $\boldsymbol{x}_p$ is the particle location (m) [12]-[13].

With pressure gradients calculated from velocity data using (1), the pressure field can be reconstructed by spatially integrating the pressure gradient field. One approach to this reconstruction calculates the pressure at each point in the flow field by integrating the pressure gradient along one path or multiple paths as:

$$\boldsymbol{p}(s) = \boldsymbol{p}(s_{ref}) + \int_{s_{ref}}^{s} \nabla \boldsymbol{p} \cdot ds, \qquad (3)$$

where $s$ is the spatial coordinate of a point in the flow field and $s_{ref}$ is a reference point with known pressure. Several path-integration algorithms have been developed [7], [14] and most employ redundant multiple path for integration, i.e., the pressure value at each point is evaluated multiple times using different paths, in order to increase accuracy. Path integration methods are rarely employed for 3D flow data due to the high computational cost. The most common approach for reconstructing pressure fields from 3D velocity data is by solving the pressure Poisson equation in the following form [8]-[10], [13], [16]:

$$\nabla^2 \boldsymbol{p} = \nabla \cdot \boldsymbol{p}_{grad,u} = -\rho \nabla \cdot (\boldsymbol{u} \cdot \nabla \boldsymbol{u}), \qquad (4)$$

where $\boldsymbol{p}_{grad,u}$ is the pressure gradient field evaluated from the velocity field and $(\nabla \cdot)$ is the divergence operator which evaluates the divergences from a vector field. Boundary conditions are required for solving (4), which can be Dirichlet boundary conditions with prescribed pressure values, Neumann boundary conditions with prescribed pressure gradient values, or a mix of the two types. As discussed in [17], both the path integration method and the method of solving the pressure Poisson equation can be regarded as global optimization formulations of the pressure-gradient spatial integration. Another method that falls into this category is a least-squares reconstruction method referred to as ordinary least-squares (OLS) reconstruction in this study [18]. For OLS, the pressure integration is performed by solving the following linear system:

$$G\boldsymbol{p} = \boldsymbol{p}_{grad,u}, \qquad (5)$$

where $G$ is the discrete gradient matrix, and $\boldsymbol{p}$ is the unknown pressure field written as a column vector. Equation (5) is an over-determined linear system for 2D and 3D flow data. The OLS method solves the pressure field by minimizing the pressure gradient residuals in a least-squares sense as:

$$p_{OLS} \equiv \underset{p}{\mathrm{argmin}}(\|\nabla \boldsymbol{p} - \boldsymbol{p}_{grad,u}\|), \qquad (6)$$

where $\| \quad \|$ is the L2 norm. In matrix form, equation (6) becomes:

$$G^T G \boldsymbol{p} = G^T \boldsymbol{p}_{grad,u} \qquad (7)$$

As stated in [19], OLS reconstruction and Poisson share the same theoretical foundation, and solving the pressure Poisson equation with Neumann boundary conditions is mathematically equivalent to the solving the OLS formulation.

Due to the measurement inaccuracies in the *in vivo* 4D-flow MRI, the calculated pressure gradient fields contain propagated errors. However, the above-mentioned pressure reconstruction methods do not have any way to account for or reduce the effect of such erroneous pressure gradient values. In order to improve the accuracy of reconstructed pressure fields, a weighted least-squares (WLS) reconstruction method for spatial integration of pressure gradients is introduced in this work. In this method, pressure fields are solved by minimizing the WLS of the pressure gradient residuals. The weights are determined based on estimated pressure gradient errors. To estimate such pressure gradient errors, velocity errors are calculated from the velocity divergence for incompressible flow and propagated through (1). Smaller weights are assigned to inaccurate pressure gradient values such that their effects are reduced during spatial integration. The performance of WLS was tested using synthetic velocity fields and *in vitro* Poiseuille flow measured using 4D-flow MRI. The method was then applied to *in vivo* 4D-flow MRI velocity data acquired for a basilar tip aneurysm and *in vitro* PTV velocity data collected in a patient-specific aneurysm model.

## II. METHODOLOGY

### A. Pressure reconstruction using weighted least-squares

Pressure gradient fields were calculated from velocity fields using (1). Velocity data employed in this study were on Cartesian grids with velocity values located on grid nodes. A second order central (SOC) difference scheme was employed to evaluate the temporal and spatial derivatives of the velocity fields. Pressure gradient values were calculated on grid nodes, then linearly interpolated to the face centers of each grid cell. The SOC scheme and grid arrangement are demonstrated in Fig. 1. SOC computes the gradient at each point from its neighboring points, e.g., $\frac{\partial p}{\partial x}(i,j) = \frac{p(i+1,j) - p(i-1,j)}{2\Delta x}$, where $\Delta x$ is the grid size. The reconstructed pressure values are on grid nodes.

The pressure field is obtained by solving

$$G^T W G \boldsymbol{p} = G^T W \boldsymbol{p}_{grad,u}, \qquad (8)$$

which gives the pressure result that minimizes the least-squares of pressure gradient residuals as

$$\boldsymbol{p}_{WLS} \equiv \underset{p}{\mathrm{argmin}}(W\|\nabla \boldsymbol{p} - \boldsymbol{p}_{grad,u}\|), \qquad (9)$$

where $W$ is the weight matrix. $W$ is a diagonal matrix containing positive elements as weights for pressure gradient values $\boldsymbol{p}_{grad,u}$. Greater weights are assigned to pressure gradient values anticipated to be more accurate. Unlike the Poisson equation, WLS reconstruction does not require boundary conditions to be explicitly assigned as the Poisson equation does. A minimum of one pressure reference point is needed. Pressure at the reference point can be obtained from direct measurement or a far-field pressure condition. If only the pressure differences between points in the flow field are of interest, the selection of reference point and reference pressure is arbitrary, e.g., zero pressure can be assigned at one point along the boundary.

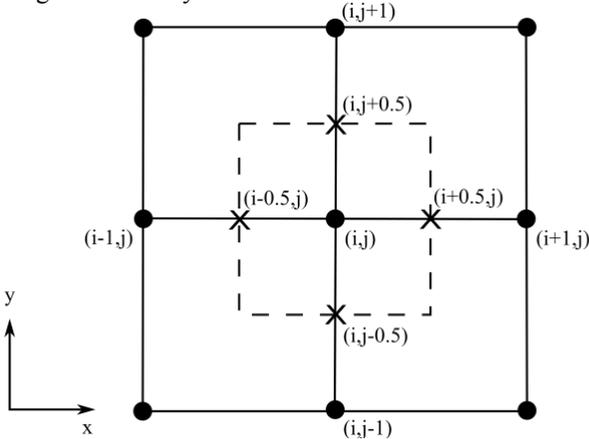

Fig. 1. Grid arrangement and SOC scheme demonstrated using a 2D Cartesian grid. The grid nodes are labeled by dots. A grid cell is drawn using dashed lines. Cell face centers are labeled by "X" marks.

### B. Velocity error estimation from spurious divergence

For incompressible flow, the divergence of the true velocity field should be zero, expressed mathematically as:
$$\nabla \cdot \boldsymbol{u_T} = 0 \tag{10}$$
where $\boldsymbol{u_T}$ is the true velocity field. Because measured velocity data inevitably contain errors, the divergence of the measured velocity field is typically nonzero. The spurious divergence equals the divergence of the velocity error field as
$$\nabla \cdot \boldsymbol{u_M} = \nabla \cdot \boldsymbol{\epsilon_u} \tag{11}$$
where $\boldsymbol{u_M}$ is the measured velocity field, $\boldsymbol{\epsilon_u}$ is the velocity error field, and $\boldsymbol{\epsilon_u} = \boldsymbol{u_M} - \boldsymbol{u_T}$. Equation (11) forms an underdetermined linear system as there are less rows than columns in the discretized divergence operator ($\nabla \cdot$). Thus, $\boldsymbol{\epsilon_u}$ cannot be uniquely determined from (11). We estimate $\boldsymbol{\epsilon_u}$ by finding the least-squares solution to (11) as
$$\widehat{\boldsymbol{\epsilon_u}} = (\nabla \cdot)^T (\nabla^2)^{-1} (\nabla \cdot \boldsymbol{u_M}) \equiv \underset{\epsilon_u}{\mathrm{argmin}}(\|\nabla \cdot \boldsymbol{\epsilon_u} - \nabla \cdot \boldsymbol{u_M}\|) \tag{12}$$
where $\widehat{\boldsymbol{\epsilon_u}}$ is the estimated velocity error field. Previous studies have similarly employed the spurious velocity divergences to estimate the uncertainty of velocity data measured using tomographic particle image velocimetry (PIV) [22].

### C. Generation of weight matrix

The pressure gradient error field is estimated by propagating $\widehat{\boldsymbol{\epsilon_u}}$ through (1) as
$$\widehat{\boldsymbol{\epsilon_{\nabla p}}} = f_{\nabla p}(\boldsymbol{u_M}) - f_{\nabla p}(\boldsymbol{u_M} - \widehat{\boldsymbol{\epsilon_u}}) \tag{13}$$
where $\widehat{\boldsymbol{\epsilon_{\nabla p}}}$ is the estimated pressure gradient error field and $f_{\nabla p}(\cdot)$ denotes evaluating (1) using the given velocity field.

Accuracy of the $\boldsymbol{p_{grad,u}}$ at each point from each time frame is determined from the weighted standard deviation (WSTD) of the estimated pressure gradient errors from neighboring points given as
$$\widehat{\sigma_{\nabla p}} = \sqrt{\frac{\sum_{i=0}^{n} w_i (\widehat{\epsilon_{\nabla p}})_i^2}{\sum_{i=0}^{n} w_i}} \tag{14}$$
where $\widehat{\sigma_{\nabla p}}$ is an estimation of the pressure gradient uncertainty and $n$ is the number of points that are employed in the WSTD calculation. Weights $w_i$ for WSTD calculations are determined using a bivariate Gaussian function
$$w_i = \exp\left(-\frac{1}{2}\left(\frac{r_t}{\delta_t}\right)^2 - \frac{1}{2}\left(\frac{r_s}{\delta_s}\right)^2\right),$$
$$\delta_t = \Delta t, \delta_s = \Delta x \tag{15}$$
where $r_t$ and $r_s$ are the spatial and temporal separation from the neighboring point to the point of interest, respectively, $\Delta t$ is the time separation between two consecutive time frames, and $\Delta x$ is the grid size. Based on the SOC scheme employed for $p_{grad,u}$ calculation, neighboring $p_{grad,u}$ values should not be correlated farther than $2\Delta x$ spatially and $\Delta t$ temporally. Thus, only points within the $r_t \leq \Delta t$ and $r_x \leq 2\Delta x$ neighborhood are employed in the WSTD calculation.

The weight matrix for WLS reconstruction is given by
$$W = diag\left(\frac{1}{\widehat{\sigma_{\nabla p}}^2}\right) \tag{16}$$
where $diag(\cdot)$ is the diagonal matrix generated from given diagonal elements. To avoid singularities due to zero weights, a lower bound of weights is given as $10^{-9}$ multiplied by the average of all weight elements.

### D. Implementation of pressure reconstruction methods

A pressure reconstruction method of solving the pressure Poisson equation (denoted as 'Poisson' herein) was employed in this study to compare to the WLS method. For the Poisson algorithm, pressure gradient fields were calculated from (1) by SOC and the grid arrangement described in section II-A. Divergence of the pressure gradients were calculated using SOC and employed as the source term for the pressure Poisson equation. Pressure gradients along the boundary of the flow field were employed as the Neumann boundary condition for solving the Poisson equation. Pressure at a boundary point was set to be zero as the reference point for both methods. SuperLU, a general-purpose library for the direct solution of large, sparse, nonsymmetric systems of linear equations [20], was employed to solve (4) and (8) for the reconstructed pressure fields.

### E. Synthetic flow fields

A 2D Lamb-Oseen vortex ring flow field was used for validating and analyzing the pressure reconstruction methods. The flow field consists of two counter-rotating vortices. Velocity of each vortex can be described by
$$u_\theta = u_{\theta max}\left(1 + \frac{1}{2\alpha}\right)\frac{r_{max}}{r}\left(1 - e^{1-\alpha\left(\frac{r}{r_{max}}\right)^2}\right). \tag{17}$$
where $u_\theta$ is the angular velocity, $r$ is the distance from the center of the vortex, $r_{max}$ is the distance where the maximum angular velocity $u_{\theta max} = 0.5 \, m/s$ is reached, and $r_{max} = \sqrt{\alpha} \times r_c$ with $r_c = 0.01 \, m$. The constant $\alpha$ was set to be 1.25643 according to [21]. The center points of vortices were separated by $2r_0$ with $r_0 = 0.01 \, m$. A free stream velocity component $u_{fs}$ was added to make the flow steady as
$$u_{fs} = u_{\theta max}\left(1 + \frac{1}{2\alpha}\right)\frac{r_{max}}{2r_0}\left(1 - e^{1-\alpha\left(\frac{r}{r_{max}}\right)^2}\right). \tag{18}$$
The exact velocity fields were generated on a uniform Cartesian grid with $65^2$ grid points. The size of the domain was $0.1 \, m \times 0.1 \, m$. The exact pressure field was obtained by numerically integrating the pressure gradients on a denser Cartesian grid with $129^2$ points. Fluid density was $1 \, kg/m^3$

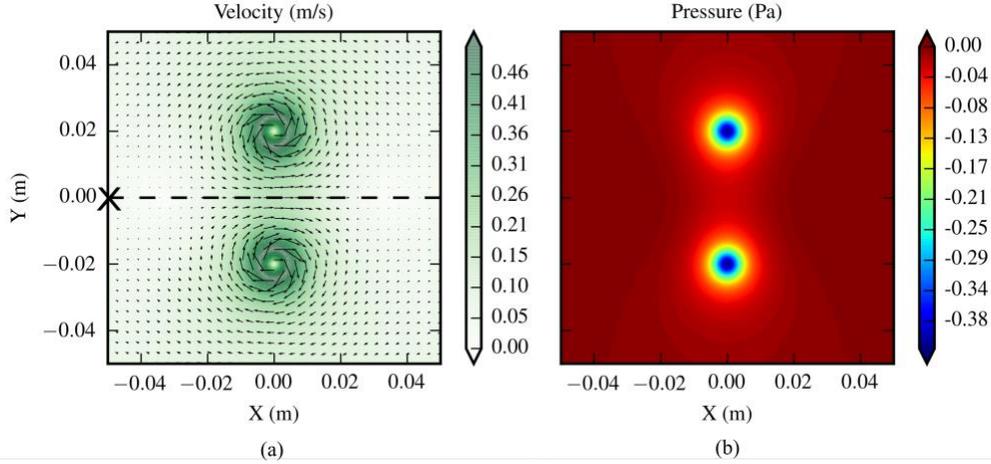

Fig. 2. Exact flow field of the 2D Lamb-Oseen vortex ring. (a) Exact velocity field. Vector indicates the flow direction and color scale of the contours corresponds to velocity magnitude. Flow field is divided by the black dashed line into top half and bottom halves. Grid point labeled by "X" is set as the reference point with zero pressure. (b) Exact pressure field.

and the flow was inviscid. Fig. 2 shows the exact velocity and pressure field.

In order to test the robustness of the pressure reconstruction methods to errors in the velocity data, noise was added to the velocity fields in a manner similar to that done in [11], [12], which is designed to mimic experimental noise. Noise was added as a vector with a normally distributed magnitude and random direction at each point. The error magnitude can be expressed by

$$\epsilon_u^i = \mathcal{N}(0, \lambda|u^i|) \tag{19}$$

where $\lambda$ is the error percentage level. Two types of velocity noise distributions were considered which are referred to as 'Uniform Noise Distribution (UND)' and 'Spatially Varying Noise Distribution (SVND)' in this study. For UND, the measurement quality was uniform across the field, and $\lambda$ was set to be consistent across the field. 26 UND test cases were generated with $\lambda$ varying linearly from 1 to 51%. For SVND, the flow field was divided into a "top half" and "bottom half" with different values of $\lambda$ applied to each half. A total of 7 cases were generated with $\lambda_{top}$ varying exponentially from 8% to 64%, and $\lambda_{bottom}$ set to 8% for all cases. For each test case with UND or SVND, 100 time frames were created with a sampling frequency of 50 Hz ($\Delta t = 0.02\ s$). 0 Pa was assigned at the left end of the dashed horizontal line in Fig. 2(a) as the reference pressure.

*F. In vitro 4D Poiseuille flow*

Experimental measurements of steady, laminar Poiseuille flow in a circular pipe were acquired using 4D-flow MRI. The Poiseuille flow allowed the usage of analytical pressure field as the benchmark to assess the accuracy of the reconstructed pressure. A blood mimicking water-glycerol solution with a density and viscosity of 1110 kg/m³ and 0.00372 Pa·s, respectively, was used as the working fluid. Gadolinium contrast (0.66 mg/mL) was added to enhance the SNR of the 4D-flow MRI scan. A computer-controlled gear pump drove the working fluid at a steady flow rate of 7.6 mL/s. The diameter of the pipe was 12.7 mm and the length was sufficiently long prior to entering the MRI field of view to ensure a fully developed velocity profile. The 4D-flow MRI scan was performed on a Siemens 3T PRISMA scanner at a spatial resolution of $0.85 \times 0.85 \times 0.8$ mm³. A total of 12 time frames were collected. The velocity encoding (venc) of this 4D-flow MRI scan (prospectively triggered time-resolved 3D PC MRI with 3-directional velocity encoding) was set to 16 cm/s, which is sufficiently high to avoid velocity wrapping. The echo time (TE) and repetition time (TR) were 5.87 ms and 8.60 ms, respectively. The yielded temporal resolution was 120.4 ms. The bandwidth was 455 kHz and flip angle was 15º. The 4D-flow MRI images were pre-processed (phase offset correction, noise filtering) using a customized Matlab-based software package, Velomap-Tool, developed at University Medical Center Freiburg [23].

*G. In vivo and in vitro flow in a basilar tip aneurysm*

*In vivo* flow data in a patient-specific basilar tip aneurysm were acquired with 4D-flow MRI on a 3T MRI scanner (Skyra, Siemens Healthcare, Erlangen, Germany) at San Francisco VA Medical Center. An ECG-gated RF spoiled 4D-flow MRI sequence (Siemens WIP sequence) was used with gadolinium contrast. Aliasing, phase offsets, and noise were corrected. Velocity data from the *in vivo* measurement was obtained on a Cartesian grid with $\Delta x = 1.25\ mm$, $\Delta y = 1.25\ mm$, and $\Delta z = 1.33\ mm$. The temporal resolution was 40.5 ms with 20 cardiac phases captured during one cycle.

*In vitro* PTV velocity data was obtained using a 1:1 scale model of the basilar tip aneurysm. To reproduce the *in vivo* flow field, the inflow was driven by a computer-controlled gear pump with the basilar inlet flow based on the *in vivo* data. DaVis 10.0 (LaVision Inc.) was used to process the particle images. Shake the Box (STB), a particle tracking method, was used to compute the velocity fields. The unstructured STB velocity fields were interpolated to a Cartesian grid with spatial resolution of 0.3 mm³ and temporal resolution of 2.5 ms. More details on the *in vivo* and *in vitro* measurement can be found in [24]. To mimic the *in vivo* 4D-flow data, another dataset was created by virtual spatial voxel averaging the *in vitro* PTV data, then temporally downsampling to the same frequency as the 4D-flow measurement. Thus, the voxel-averaged and subsampled dataset (referred to as 'voxavg' herein) had the same spatial and temporal resolution as the *in vivo* 4D-flow MRI dataset.

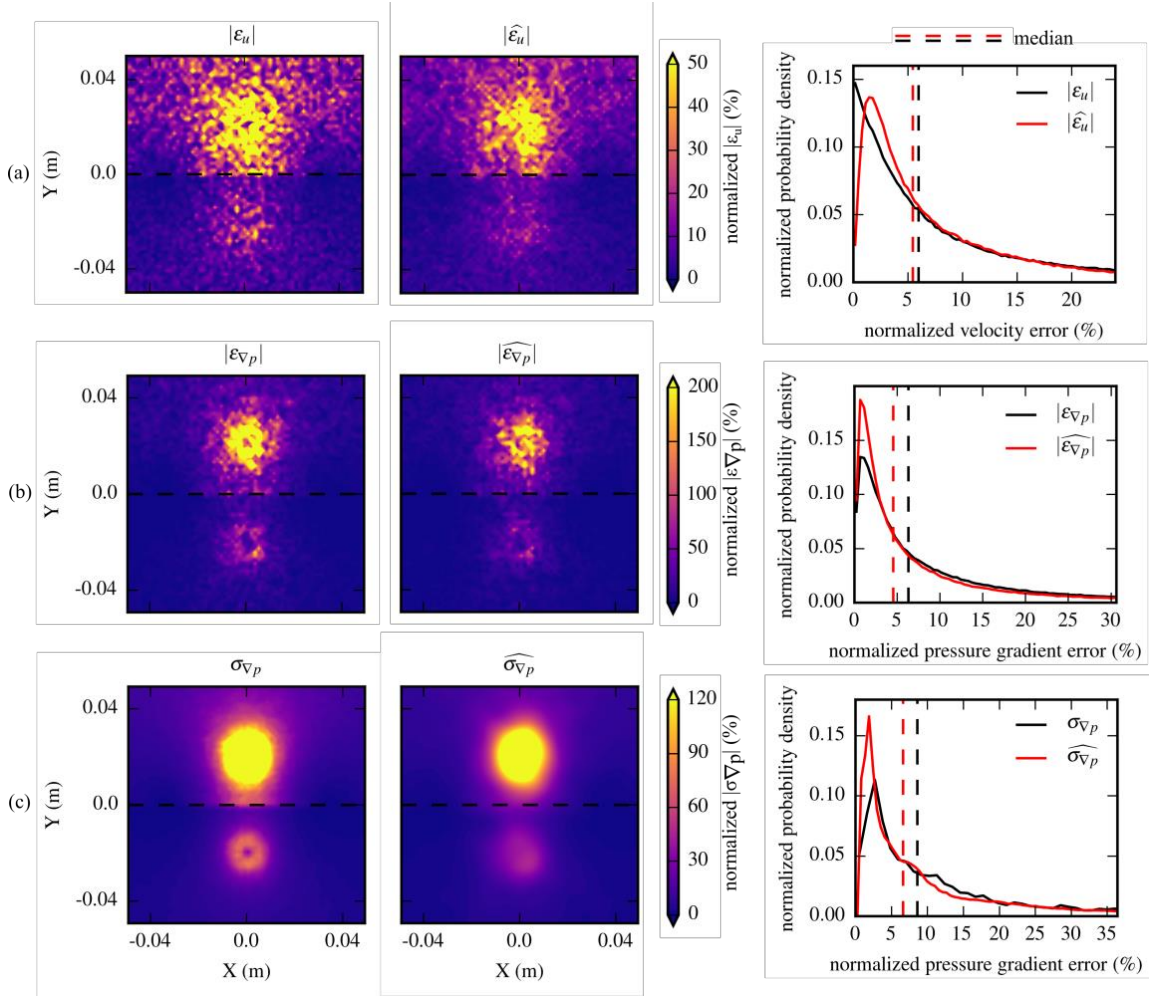

Fig. 3 Examples of the estimated error distributions compared with the exact error distributions from the SVND case with $\lambda_{top}$ being and $\lambda_{bottom}$ being 8%. The first two columns are the spatial distributions. The last column shows histograms of error magnitudes. The dashed vertical lines represent the medians of the distributions. (a) Comparison between exact velocity error magnitudes and estimated velocity error magnitudes. (b) Comparison between exact pressure gradient error magnitudes and estimated pressure gradient error magnitudes. (c) Comparison between the pressure gradient uncertainties and the WSTD of the estimated pressure gradient errors.

## III. RESULTS

### A. Lamb-Oseen vortex ring

*1) Velocity and pressure gradient error estimation*

To validate the error estimation algorithm employed in this study, the estimated velocity and pressure gradient errors were compared with the exact errors from all the Lamb-Oseen vortex cases. As a demonstration, the distributions of estimated and exact errors from the case with $\lambda_{bottom}$ and $\lambda_{top}$ being 8 and 32%, respectively, are shown in Fig. 3. Fig. 3(a) and (b) present the comparisons on velocity error magnitudes ($|\widehat{\epsilon_u}|$ versus $|\epsilon_u|$) and pressure gradient error magnitudes ($|\widehat{\epsilon_{\nabla p}}|$ versus $|\epsilon_{\nabla p}|$), respectively. Fig. 3(c) compares $\widehat{\sigma_{\nabla p}}$ with the pressure gradient uncertainty ($\sigma_{\nabla p}$) evaluated as the root-mean-square (RMS) of $\epsilon_{\nabla p}$ from all time frames. For both estimated and exact errors, the magnitudes were greater in the top half of the field than in the bottom half, and greater in the vortices than in the ambient regions, as suggested by the spatial distributions. The estimated magnitudes were lower than the corresponding exact magnitudes as suggested by the medians from the histograms in Fig. 3. The median of $|\widehat{\epsilon_u}|$ was 5.5% while it was 6.0% for $|\epsilon_u|$. The medians of $|\widehat{\epsilon_{\nabla p}}|$ and $|\epsilon_{\nabla p}|$ were 4.5% and 6.3%, respectively. The median of $\widehat{\sigma_{\nabla p}}$ was 6.6% while it was 8.6% for $\sigma_{\nabla p}$. The error estimation algorithm performed consistently for all the cases.

*2) Pressure error analysis*

The errors in the pressure fields reconstructed using Poisson and WLS were analyzed and compared. Pressure errors ($\epsilon_p$) were quantified as the deviation between the reconstructed pressure and the exact pressure. $\epsilon_p$ and $\epsilon_u$ were normalized by the RMS norm of the exact pressure field and velocity field, respectively. The distributions of velocity and pressure error magnitudes are shown in Fig. 4 for three cases with $\lambda_{bottom}$ being 8% while $\lambda_{top}$ being 8%, 32%, and 64%. The spatial distributions in Fig. 4 presents the RMS of normalized errors from all time frames. As suggested by both the spatial distributions and the histograms, the pressure error magnitudes for WLS ($|\epsilon_{p,WLS}|$) were lower than those for Poisson ($|\epsilon_{p,Poisson}|$). The medians of $|\epsilon_{p,WLS}|$ and $|\epsilon_{p,Poisson}|$ were 0.8% and 1.2%, respectively, with $\lambda_{top}$ being 8%, 1.6% and 3.6% with $\lambda_{top}$ being 32%, and 2.4% and 8.3% with $\lambda_{top}$ being 64%.

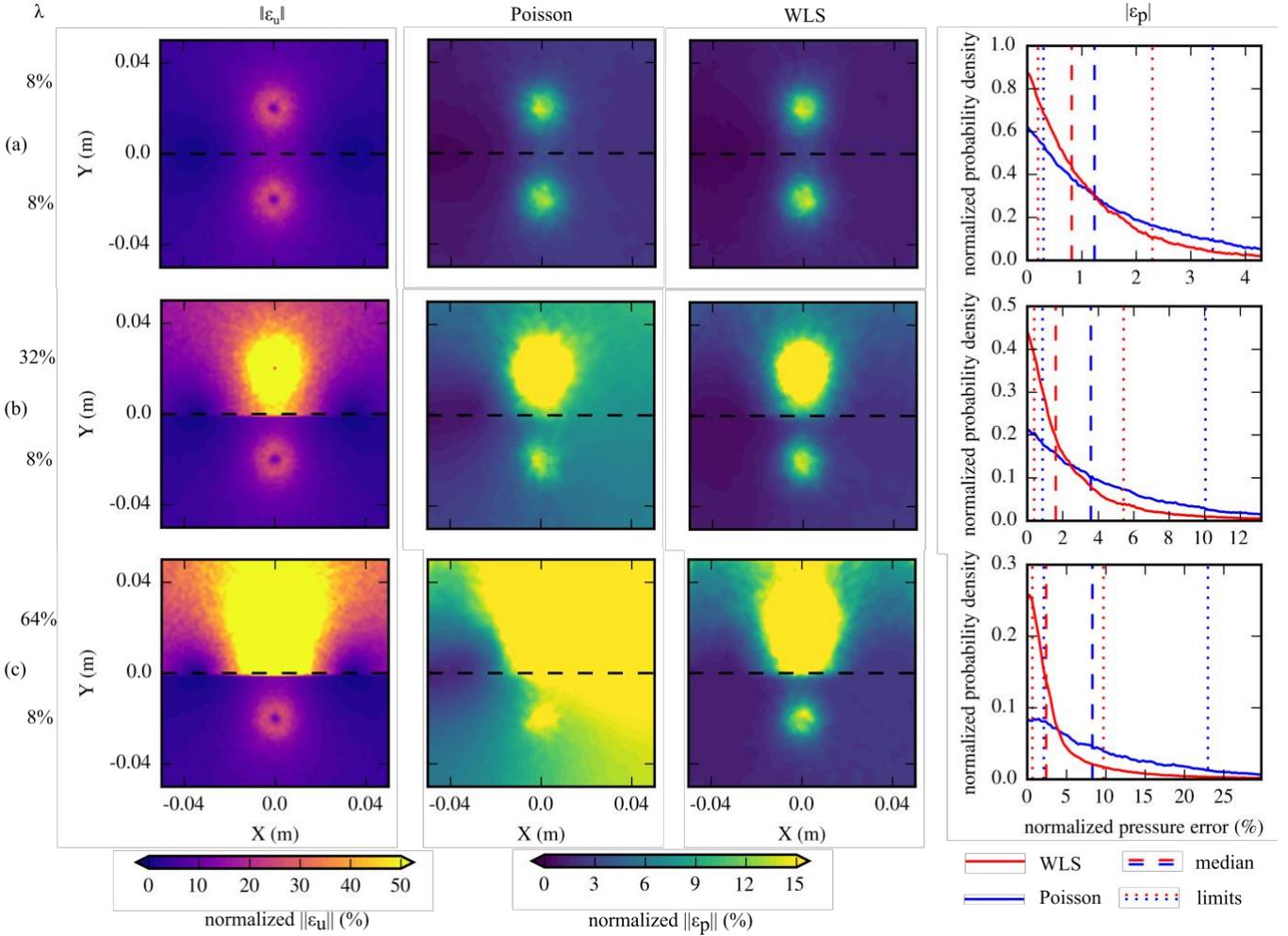

Fig. 4 The spatial distributions of normalized velocity error magnitudes (1st column), normalized pressure error magnitudes (2nd and 3rd columns), and the histograms of normalized pressure error magnitudes (last column) for three test cases of the 2D synthetic flow. The errors were normalized by the RMS of the exact fields. The vertical dashed lines in the histograms are medians of the distributions. The vertical dotted lines are the lower and upper limits of pressure error magnitudes. (a) $\lambda_{top} = 8\%$, $\lambda_{bottom} = 8\%$. (b) $\lambda_{top} = 32\%$, $\lambda_{bottom} = 8\%$. (c) $\lambda_{top} = 64\%$, $\lambda_{bottom} = 8\%$.

The performances of the pressure reconstruction methods are compared in Fig. 5(a) using results from all test cases with UND. The velocity error level for each case was determined as the median of the normalized velocity error magnitudes. As $\lambda$ changed from 1 to 51%, the velocity error ranged from 0.39 to 19.9%. Similarly, the pressure error levels were determined as the median of the normalized pressure error magnitudes. For the noise level range used here, the pressure error for WLS increased from 0.10 to 6.0%, while it increased from 0.15 to 9.9% for Poisson. Thus, WLS maintained a 50% improvement on median pressure error over Poisson. Additionally, the lower and upper limits of the pressure errors were given as the 15.75th and 84.25th percentiles of the absolute error distribution, respectively. The upper limit for WLS increased from 0.28 to 17.3%, while it increased from 0.40 to 27.3% for Poisson. The lower limit for WLS increased from 0.025 to 1.44%, while it increased from 0.037 to 2.39% for Poisson.

Fig. 5(b) compares the error levels for the two methods from all cases with SVND. As $\lambda_{top}$ changed from 8 to 64% and $\lambda_{bottom}$ stayed at 8%, the overall velocity error level increased from 3.1 to 7.9%. The pressure error for WLS increased from 0.8 to 2.4%, while the it increased from 1.2 to 8.3% for Poisson. The lower error limit ranged from 0.2 to 0.6% for WLS and from 0.3 to 2.1% for Poisson. The upper limit ranged from 2.4 to 9.7% for WLS and from 3.4 to 23.0% for Poisson. In addition to the overall pressure error level, the pressure error level within each half of the field was quantified and presented in Fig. 5(c) and 5(d), respectively. The pressure error level in the top half ranged from 0.8 to 4.9% for WLS and from 1.2 to 11.4% for Poisson, while that in the bottom half ranged from 0.8 to 1.4% for WLS and from 1.2% to 6.4% for Poisson.

### B. In vivo 4D Poiseuille flow

The analytical velocity field of the Poiseuille flow is given by:

$$W = -\frac{1}{4\mu}\frac{dP}{dz}(R^2 - r^2), \tag{20}$$

where $W$ is the axial (along z-axis) velocity component (m/s), $r$ is the radial distance from the pipe centerline (m) which equals to $\sqrt{x^2 + y^2}$ and $\frac{dP}{dz}$ is the axial pressure gradient (Pa/m). The velocity components along other axes ($U$ and $V$) are 0. The axial pressure gradient is defined by:

$$\frac{dP}{dz} = -\frac{8\mu Q}{\pi R^4}, \tag{21}$$

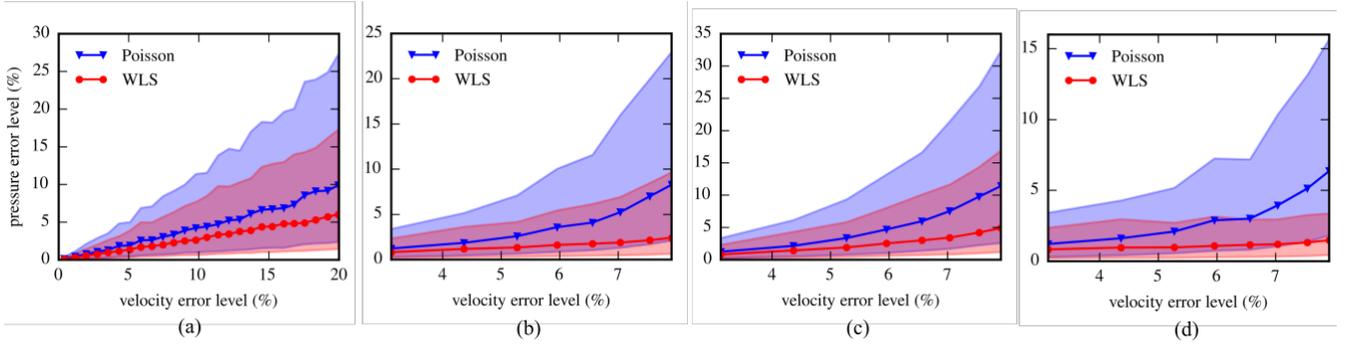

Fig. 6 The pressure error level versus velocity error level from the test cases of synthetic flow. The error levels were determined as the medians of error magnitudes. The shaded areas are bounded by the upper and lower limits of pressure error magnitudes. (a) Results from UND cases with $\lambda$ changing from 1% to 51%. (b) Results from SVND cases with $\lambda_{top}$ changing from 8% to 64% and $\lambda_{bottom}$ being 8%. (c) Pressure error levels in the top half of the flow fields shown as a function of velocity error levels for SVND cases. (d) Pressure error levels in the bottom half of the flow fields for SVND cases.

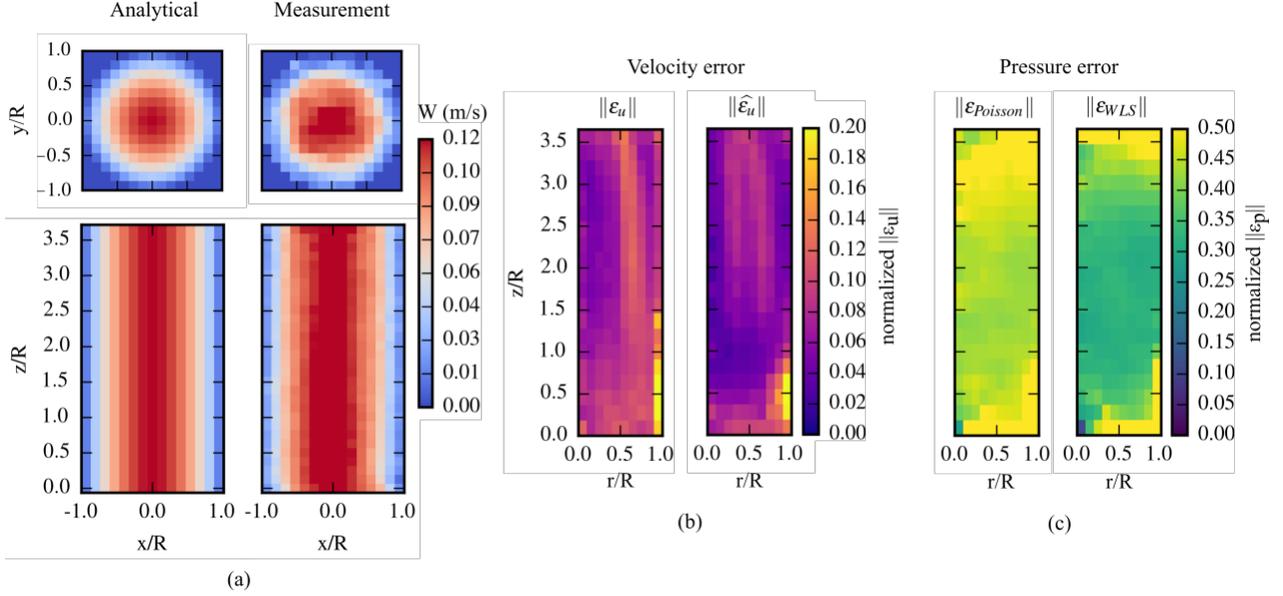

Fig. 5 (a) The velocity profiles of laminar pipe flow from analytical solution and measurement. The velocity profiles are shown on x-y plane at z=0 mm and on x-z plane at y=0 mm. (b) The spatial distributions of normalized velocity errors shown as functions of r and z. (c) The spatial distributions of normalized pressure errors for Poisson and WLS.

where $Q$ is the volumetric flow rate (m³/s). This yields a linear analytical pressure drop along the pipe. The analytical velocity field is shown in Fig. 6(a) together with a time frame from the measured velocity data. Fig. 6(b) compares the velocity errors ($\epsilon_u$) evaluated as the deviations between the analytical velocity and measured velocity with the velocity errors estimated based on velocity divergence ($\widehat{\epsilon_u}$). The error magnitudes were normalized by the centerline velocity of the analytical field. For both $\epsilon_u$ and $\widehat{\epsilon_u}$, the magnitudes were greater near the wall or close to the ends of the pipe.

Instantaneous pressure fields were reconstructed using Poisson and WLS from the measured velocity fields. The origin (r=0 mm, z=0 mm) was selected as the reference point with zero pressure. The pressure errors were evaluated as the deviation between analytical pressure and reconstructed pressure, then normalized by the analytical pressure drop across the measurement region ($\Delta p_{analytical}$). Spatial distributions of the normalized pressure error magnitudes are presented as functions of r and z in Fig. 6(c). The pressure in the middle region of the pipe had significantly lower error when using WLS. To confirm this notion, the histograms of the relative pressure error magnitudes are shown in Fig. 7. The median of pressure error magnitude was 24.6 % for WLS and 35.6% for Poisson. The lower error limit was 7.8 % for WLS and 11.2% for Poisson. The upper error limit was 53.1% for WLS and 64.5% for Poisson.

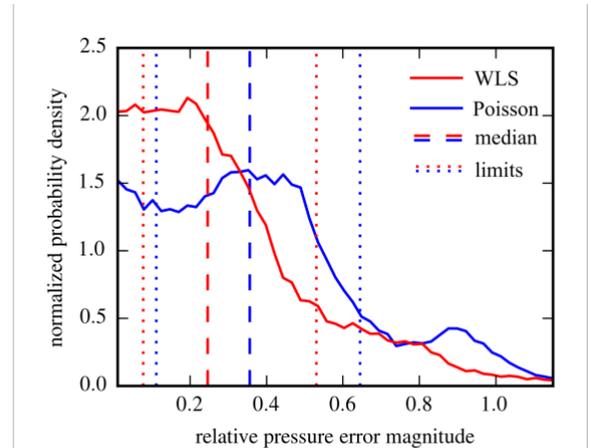

Fig. 7 Histograms of the pressure error magnitudes from the pressure fields reconstructed using Poisson and WLS.

*C. Flow in basilar tip aneurysm*

The velocity pathlines at peak systole are presented in Fig. 8 for the *in vivo* 4D-flow and *in vitro* PTV data. The flow structures of both datasets are consistent. For both datasets, the inflow comes from the basilar artery and forms a vortical structure in the aneurysmal sac. However, the PTV data was obtained with higher spatial and temporal resolution and was contaminated with less noise. The average flow rate error (difference between inflow and outflow flow rates) was 24.0% for the 4D-flow data, and 6.9% for the PTV data, suggesting better accuracy for the PTV data.

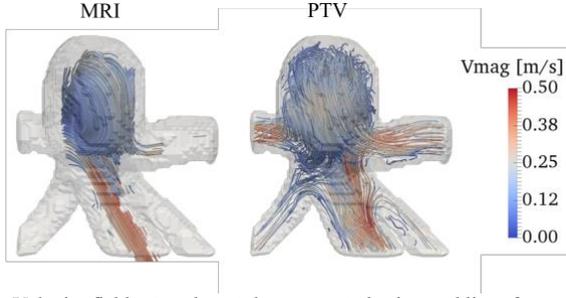

Fig. 8 Velocity fields at peak systole represented using pathlines from *in vivo* 4D-flow MRI and *in vitro* PTV measurements. Shaded regions represent the geometry of the aneurysm.

As the flow structures are consistent, similar pressure fields are anticipated for the *in vivo* data and *in vitro* data. Thus, even given the noisy and under-resolved *in vivo* 4D-flow data, a robust pressure reconstruction method should be capable of obtaining similar pressure fields to those reconstructed from the *in vitro* PTV data. Fig. 9 compares the pressure reconstruction using the Poisson and WLS methods using the *in vivo* 4D-flow data and *in vitro* PTV and PTV voxavg data at two different planes at peak systole. The pressure near the wall and inside arteries was not calculated from *in vivo* data due to insufficient spatial resolution such that the spatial derivatives cannot be resolved. Thus, the pressure in these regions was not included in the following comparisons with pressure calculated from *in vitro* data. The pressure fields were normalized by the maximum pressure difference within the aneurysmal sac ($\Delta p_{max}$) from each modality. Additionally, the histograms of the reconstructed pressure values in the aneurysmal sac from the entire cardiac cycle are shown in Fig. 9(d). WLS showed better agreement across all datasets than Poisson in both the spatial distributions as well as the pressure histograms. Given in Fig. 9(d), the median pressure values by Poisson were −0.4 Pa for 4D-flow, −8.9 Pa for voxavg, and −7.1 Pa from PTV. The median values obtained using WLS were −4.1 Pa for 4D-flow, −4.7 Pa for voxavg, and −5.5 Pa for PTV. The standard deviation of the medians was 3.7 Pa for Poisson and 0.6 Pa for WLS. This indicates that WLS maintained a tighter spread of the pressure values and more similarity across modalities, suggesting it is more robust to low-resolutions and high-noise velocity fields. From each modality, the deviations between the pressure fields reconstructed by WLS and the pressure fields reconstructed by Poisson were quantified and normalized by $\Delta p_{max}$. The total RMS of the normalized deviations was defined as the "effectiveness" of WLS on improving pressure reconstruction for each modality. The effectiveness was 28.7% for 4D-flow, 17.9% for voxavg, and 8.7 % for PTV.

IV. DISCUSSION AND CONCLUSIONS

In this study we introduced a method which uses weighted least-squares for pressure integration. By assigning lower weights to less accurate velocity measurements and thus pressure gradient values, the WLS method reduces the effects of noisy measurements during the spatial integration, and improves the accuracy of the reconstructed pressure. Poisson and OLS can be seen as particular cases of WLS with uniform weights assigned to the pressure gradients. The accuracy of WLS relies on proper weight assignment. In this study, the weights were informed by the estimated velocity errors based on velocity divergence. Comparisons between exact velocity error and estimated velocity error demonstrated that the velocity error estimation algorithm used here was capable of recognizing high-error regions such that lower weights were assigned to the less accurate pressure gradients in these regions. Although the velocity and pressure gradient error magnitudes were found to be slightly underestimated by this algorithm, this is not expected to affect the performance of WLS. This is because underestimating the error magnitudes would have a similar effect as normalizing the weights by a constant greater than 1. Further, the weight matrix W appears on both sides of (8), therefore the weights can be normalized by any arbitrary nonzero, real constant while the pressure results remain the same. Thus, the spatial distribution of the estimated error is primarily what effects the accuracy of WLS as opposed to the error values themselves. It should also be noted that the weights can be informed by the pressure gradient reliabilities estimated using other algorithms. For velocity fields measured using PIV, there are algorithms to estimate the spatial distributions of velocity uncertainties [25],[26] and the pressure gradient uncertainties [27]. However, a corresponding algorithm for 4D-flow data has not been developed. The divergence-based algorithm employed in this study can be applied to velocity data measured from incompressible flows regardless of the measurement modality.

The WLS method reduces the spatial propagation of errors during pressure integration. From the spatial distributions of pressure errors for the synthetic Lamb-Oseen vortex flow in Fig. 4, it can be observed that WLS reduced pressure errors in the ambient regions as the greater errors were more confined to the vortices. In addition, $|\epsilon_{p,WLS}|$ in the bottom half of SVND cases was significantly less affected by the increase of $\lambda_{top}$ as compared with $|\epsilon_{p,Poisson}|$. As observed in Fig. 5(d), as $\lambda_{top}$ increased from 8% to 64% and $\lambda_{bottom}$ stayed at 8%, the increase of the pressure error in the bottom half was 433% for Poison while only 75% for WLS. This is also suggested by the spatial distributions of pressure errors from *in vitro* Poiseuille flow in Fig. 6(b). WLS confined the pressure errors to the regions with greater velocity errors (near the ends of the pipe) compared with Poisson. In previous studies, the spatial error propagation was reduced by segmenting the flow field into subdomains based on local velocity reliability, then reconstructing the pressure field in each subdomain sequentially in a descending order of reliability [18]. However, such an algorithm requires that the different levels of measurement reliability are spatially separable in the flow field such that the subdomains can be properly segmented. The WLS method proposed here does not require any segmentation, making it more usable across a larger variety of flow fields.

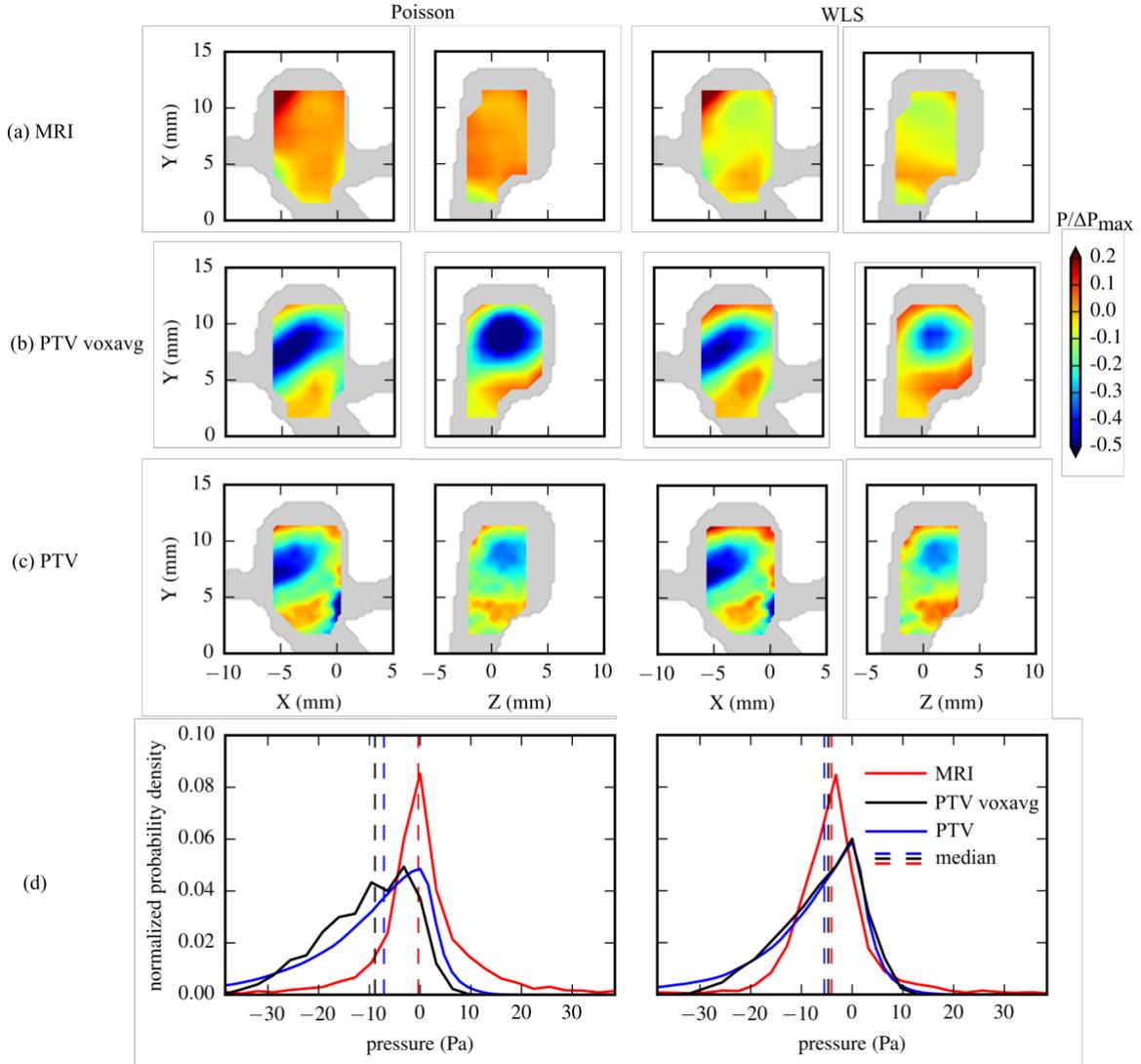

Fig. 9. Spatial and probability density distributions of pressure fields reconstructed using Poisson and WLS from each dataset. The spatial distributions are presented by the normalized pressure values on a x-y plane and a y-z plane cutting through the aneurysm sac at peak systole. Shaded regions correspond to the geometry of the aneurysm. (a) Spatial distributions for *in vivo* 4D-flow MRI data. (b) Spatial distributions for voxel-averaged (voxavg) PTV data. (c) Spatial distributions for *in vitro* PTV data. (d) Probability density distributions of the reconstructed pressure values within the aneurysmal sac.

Improvement on pressure accuracy by WLS was more significant for velocity data with a greater range of errors. Results from synthetic Lamb-Oseen vortex flow fields demonstrated that the improvement by WLS was more significant for SVND cases with greater $\lambda_{top}$. Given in Fig. 5(b), the pressure error level for Poisson was 240% larger than that for WLS with $\lambda_{top}$ of 64%, and 50% when $\lambda_{top}$ was 8%. This was also reflected by the results from the aneurysm flow. Among the three datasets, the *in vivo* 4D-flow data contained the widest range of velocity errors. The pressure fields reconstructed from 4D-flow data using WLS were more consistent with the observed flow structure compared with Poisson as suggested by Fig. 9(a). Specifically, the center of the aneurysmal sac was expected to be a low-pressure region given the vortical flow in that region, and the high-pressure regions were expected to be near the inlet and the tip of the aneurysmal sac based on the flow deceleration. These anticipated distributions were observed using WLS, but not using Poisson. However, the pressure fields reconstructed from the *in vitro* datasets using the two methods were all consistent with the expected pressure distribution. The corresponding effectiveness of WLS was highest (28.7%) for 4D-flow data compared with other datasets (17.9% for voxavg and 8.7% for PTV). Overall, the analyses here suggest that WLS improved the pressure reconstruction from less accurate velocity data as compared to the Poisson method.

A limitation of this study was that no benchmark pressure was available for the comparison between the pressure fields reconstructed from the aneurysm data such that errors in the reconstructed pressure fields could not be quantified. Instead, we could only compare the pressure fields calculated from *in vivo* data and *in vitro* data based on the notion that the pressure fields should be similar as the flow structures are consistent. Although the *in vivo* and *in vitro* flow data were found to be in good agreement [24], they were not exactly the same and thus the pressure fields could maintain inherent differences. A comparison of the reconstructed pressure to a direct pressure measurement would improve the evaluation of the WLS pressure accuracy and should be explored in future work.

There are also several limitations of the WLS pressure reconstruction method. The error estimation algorithm employed in this study can only be applied to incompressible flows as the divergence-free assumption is invalid for compressible flows. In addition, the algorithms for error estimation and pressure gradient calculation are only applicable to velocity data which fully resolves the gradients along all dimensions. For 3D flows, volumetric data with all 3 velocity-components are required. 2D planar velocity data or 3 velocity components captured on a 2D plane measured from 3D flow would not be sufficient because the velocity gradient perpendicular to the measurement plane is not resolvable. However, this algorithm can be applied to 2D planar data if the flow is uniform along the perpendicular dimension, such as the 2D synthetic flow employed in this study. Another limitation of WLS is that the velocity data needs to be temporally and spatially resolved to ensure accurate derivative evaluation. However, this is a limitation for most pressure reconstruction methods.